# Enhancement in surface mobility and quantum transport of $Bi_{2-x}Sb_xTe_{3-y}Se_y$ topological insulator by controlling the crystal growth conditions

Kyu-Bum Han[1,†], Su Kong Chong[2,†], Akira Nagaoka[3], Suzanne Petryk[4], Michael A Scarpulla[1,3], Vikram V. Deshpande[2], and Taylor D. Sparks[1,*]

[1]Department of Materials Science and Engineering, University of Utah, Salt Lake City, Utah 84112 USA

[2]Department of Physics and Astronomy, University of Utah, Salt Lake City, Utah 84112 USA

[3]Department of Electrical Engineering, University of Utah, Salt Lake City, Utah 84112 USA

[4]Department of Computer Science, Cornell University, 402 Gates Hall, Ithaca, NY 14853, USA

[†]These authors contributed equally.

*Corresponding author: sparks@eng.utah.edu

## ABSTRACT

**Despite numerous studies on three-dimensional topological insulators (3D TIs), the controlled growth of high quality (bulk-insulating and high mobility) TIs remains a challenging subject. This study investigates the role of growth methods on the synthesis of single crystal stoichiometric $BiSbTeSe_2$ (BSTS). Three types of BSTS samples are prepared using three different methods, namely melting growth (MG), Bridgman growth (BG) and two-step melting-Bridgman growth (MBG). Our results show that the crystal quality of the BSTS depend strongly on the growth method. Crystal structure and composition analyses suggest a better homogeneity and highly-ordered crystal structure in BSTS grown by MBG method. This correlates well to sample electrical transport properties, where a substantial improvement in surface mobility is observed in MBG BSTS devices. The enhancement in**

**crystal quality and mobility allow the observation of well-developed quantum Hall effect at low magnetic field.**

Three-dimensional topological insulator (3D TI) is a novel state of quantum matter which shows potential applications in spintronics[1,2] and quantum computing[3,4] owing to the unique spin-momentum locked surface states. Charge transport in such Dirac dispersion surface states is less sensitive than ordinary conductive materials to defects because the surface states are protected by time-reversal symmetry[5]. The surfaces states in 3D TI have been extensively studied and confirmed using angle-resolved photoemission spectroscopy[6,7], scanning tunneling microscopy[8] and electric transport measurements[9,10]. However, the typical strong 3D TIs, $Bi_2Se_3$ and $Bi_2Te_3$ suffer from their deep bulk-doping[11]. The difficulty in isolating bulk contributions from surface state transport limits the implementation of TIs in electronic devices[12]. More recent studies focus on the ternary and quaternary tetradymite Bi-based TI, namely $Bi_2Te_2Se$ (BTS)[13-17] and $Bi_{2-x}Sb_xTe_{3-y}Se_y$[18-25], which show a large bulk resistivity, ambipolar surface transport and clear quantum oscillations. Particularly, the stoichiometric $BiSbTeSe_2$ (BSTS) is a very promising TI candidate as it manifests high quality integer quantum Hall effect originating from its surface states[23-25].

Various growth techniques have been reported to prepare BSTS single crystals. Among those, melting[17-22,24,25] and vertical Bridgman[13,14,23] methods are the most widely used synthetic methods. The single crystal growth methodology for melting method (as illustrated in Fig. 1(a)) generally involves the following procedures: (i) Melting and mixing the source materials in a sealed ampoule at a temperature above the melting temperature. (ii) The molten materials are annealed at a temperature just above the melting point for a long period to allow atomic diffusion and nucleation of a crystalline lattice. (iii) The materials are cooled down to the room temperature at a very slow rate to minimize additional nucleation sites in favor of growing the existing single crystal. Obviously, melting growth can cause polycrystalline ingot as there is no fixed single nucleation site. However, numerous literature has claimed that they achieved single crystal growth using the melting method[17-22,24,25]. On the other hand, the vertical Bridgman method applies a vertical motion to translate the materials along a temperature gradient from hot to cold zones at a very slow rate, as shown in Fig. 1(b). The source materials are gradually melted and solidified from the bottom to the top of an ampoule in order to yield single crystal growth. The crystal quality greatly depends on the translation rate as the slow solidification encourages the sample homogeneity and minimizes the crystal defects[26].

Control of stoichiometry of the $Bi_{2-x}Sb_xTe_{3-y}Se_y$ is crucial to control the chemical potential of the surface states to be within the bulk gap[19]. Despite many studies, the single crystal growth of stoichiometric BSTS with remains challenging. Two processing conditions have not been fully explored in BSTS growth which may play an important role in crystal quality: (i) Rather than melting elemental precursors, it is possible to pre-react precursors to form a more homogenous compound prior to single crystal growth. (ii) Limiting the empty ampoule headspace. For the BSTS system, the starting materials have very different melting temperatures ($T_m$) ranging from 220°C (Se) to 631°C (Sb). Therefore, an initial mixing step is crucial to yield highly stoichiometric and homogeneous BSTS single crystals. In the mixing process, the starting materials are melted and stirred multiple times over a period at high temperature of 850°C to obtain the homogeneity[18-20]. For the second aspect (suitable only to vertical Bridgman growth), a quartz rod is inserted at the top of the BSTS polycrystalline ingot to reduce the space for evaporating materials, and thus minimize the loss of low $T_m$ elements (refer to schematic in Fig.1(b)). As the first and second aspects can only be applied separately to the melting and Bridgman growth methods, it is practical to investigate a two-step growth technique by combining melting and Bridgman methods.

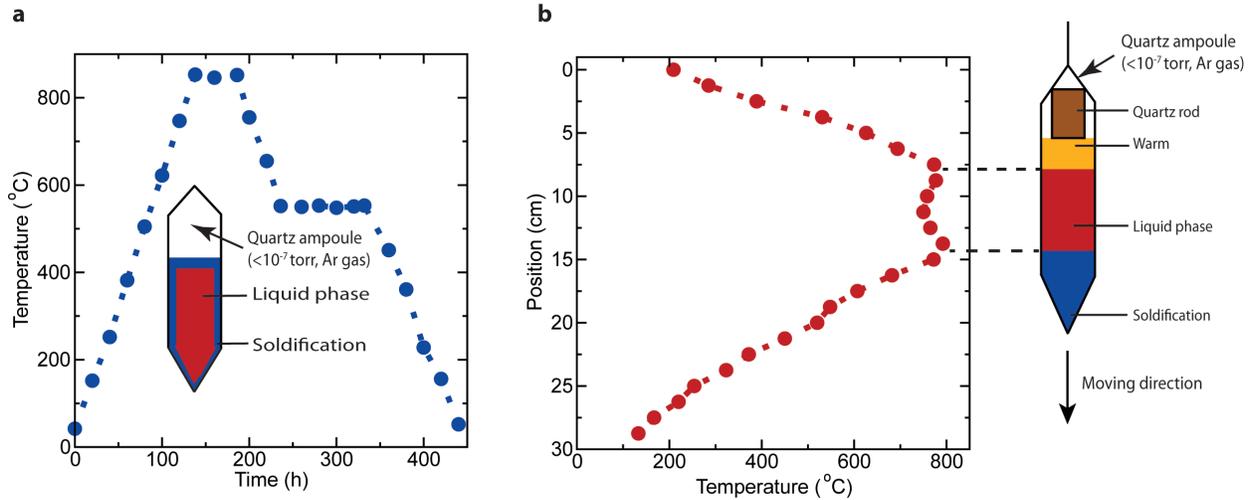

**Figure 1.** The measured temperature profiles of the melting (a) and vertical Bridgman (b) furnaces. The schematics of ampoules in (a) and (b) illustrate the crystal growth mechanism for the respective methods.

In this work, we provide a systematic study on the crystal growth of BSTS using three different methods, namely melting growth (MG), vertical Bridgman growth (BG), and two-step melting-Bridgman growth (MBG). The MG and BG parameters are illustrated in Fig. 1(a) and (b), respectively. MBG begins with growing an ingot via MG (possibly single crystal), and then follows with recrystallization into a single crystal by BG. The single crystal BSTS samples grown by MG, BG and MBG methods are shown in Fig. S1. The composition homogeneity, crystal structure and low temperature magnetoelectric transport of the three types of BSTS samples are characterized. We show that the structural and electrical transport properties depend strongly on the growth method. The crystallinity of the BSTS, identified by structural properties, is correlated to the surface mobility and the quantum transport properties.

## Results

**Stoichiometry study.** The elemental composition of the as-grown BSTS samples was investigated using both energy-dispersive X-ray spectroscopy (EDS) and inductively coupled plasma mass spectrometry (ICP-MS). The results are presented in Table 1. The EDS data were taken from the homogeneous part of the crystals (excluding the side surfaces). The EDS results are presented in the form of $Bi_{2-x}Sb_xTe_{3-y}Se_y$ converted from the atomic percentage of the quantitative data. The BSTS crystals grown by both MG and BG methods showed similar composition ratios which differed from MBG method. The composition of the MBG sample was found to be ~10% closer to the stoichiometric ratio (1:1:1:2 for Bi:Sb:Te:Se) compared to the MG and BG samples. To further confirm the composition results, we utilized the more accurate ICP-MS for the elemental characterization. Consistent with the EDS analysis, both MG and BG samples showed similar composition in the ICP-MS spectrum and MBG sample was again found to more closely match the desired stoichiometric $BiSbTeSe_2$.

**Table 1.** The elemental composition measured using EDS and ICP-MS for different methods prepared BSTS samples.

| Growth method | EDS ($\pm 2\%$) | ICP-MS ($\pm 3\%$) |
|:---:|:---:|:---:|
| MG | $Bi_{1.00}Sb_{0.90}Te_{0.95}Se_{2.15}$ | $Bi_{0.75}Sb_{1.25}Te_{0.56}Se_{2.44}$ |
| BG | $Bi_{0.92}Sb_{0.95}Te_{0.98}Se_{2.15}$ | $Bi_{0.79}Sb_{1.21}Te_{0.62}Se_{2.38}$ |
| MBG | $Bi_{0.96}Sb_{1.02}Te_{0.97}Se_{2.05}$ | $Bi_{0.89}Sb_{1.11}Te_{0.88}Se_{2.12}$ |

**Crystallinity evaluation.** The crystal structure and crystal quality of the BSTS samples were studied by X-ray diffraction (XRD). Fig. 2(a) shows the XRD patterns collected from the (00*l*) crystal surface. The crystallographic planes of the corresponding diffraction peaks were well-indexed in the figure with no evidence of polycrystalline grains. The highest intensity diffraction peak, (006), for the BSTS samples grown by different methods is compared in Figure 2(b). The MG BSTS showed a broad diffraction peak with two shoulders at both sides of the peak. The shoulder broadening could be due to the inhomogeneous compositions resulting in anti-site or vacancy defects in the crystals[14,18]. The diffraction peak of BG BSTS revealed only high angle broadening, suggesting a more homogeneous composition due to the zone melting and solidification process in the Bridgman method. Similar observations had been found for the vertical Bridgman growth in $Bi_2(Te_{1-x}Se_x)_3$ system[14]. The MBG BSTS showed a more symmetric Gaussian curve and narrower peak width, which indicated a better crystallinity and homogeneity compared to the MG and BG BSTS. For a quantitative comparison, the full-width at half maximums (FWHM) of the (006) diffraction peaks for the three BSTS growth methods are plotted in Figure 2(c). The smaller FWHM generally implied the improvement in crystal lattice arrangement as observed in MBG BSTS.

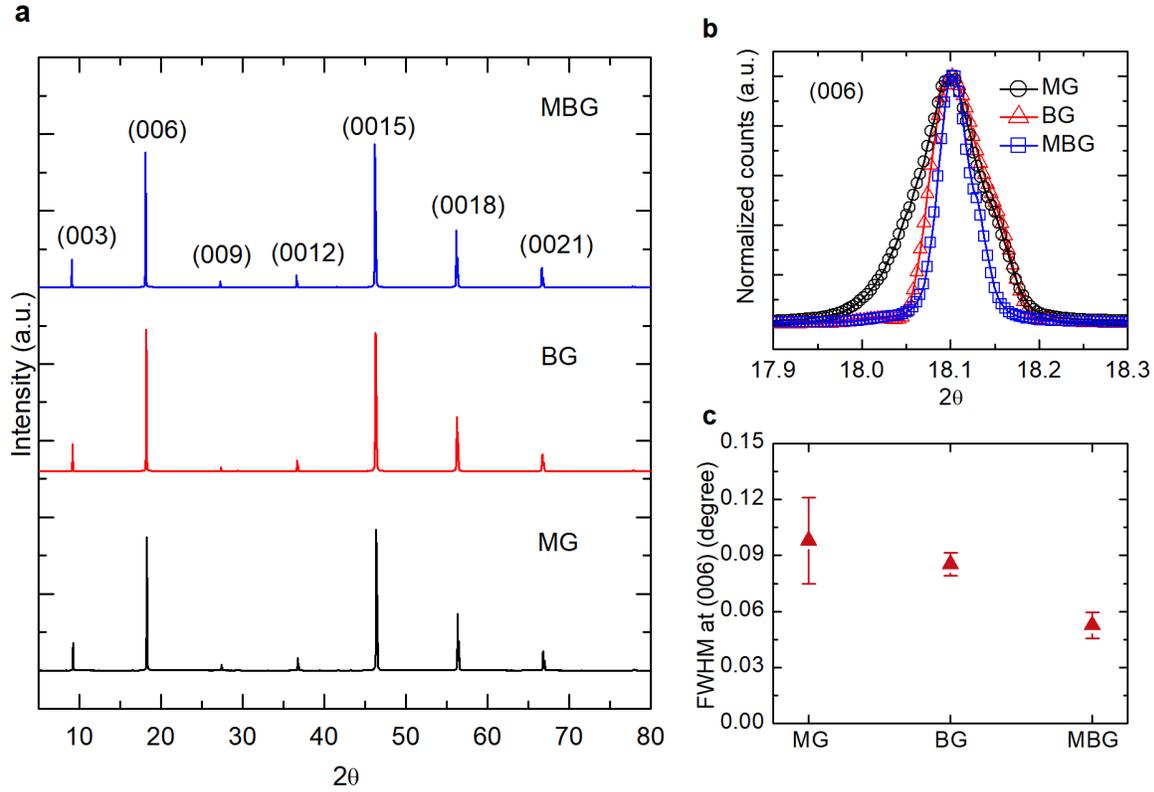

**Figure 2.** XRD patterns scanned at (00*l*) crystal surface for BSTS grown by MG, BG, and MBG methods (a). The (006) diffraction peaks (b) and their FWHMs (c) for the three BSTS growth methods.

**Crystal structure investigation.** The crystal structures of the BSTS were examined from the powder XRD patterns as presented in Figure 3(a). The Rietveld refinements from the diffraction data were well-matched to the rhombohedral structure in space group of $R\bar{3}m$. This agreed with the reported XRD data for BSTS topological insulators[18,20]. The lattice parameters of the MG BSTS ($a$= 4.1021(7) Å and $c$= 29.0822(5) Å), BG BSTS ($a$= 4.1240(3) Å and $c$= 29.1209(7) Å), and MBG BSTS ($a$= 4.1623(6) Å and $c$= 29.2019(4) Å) were determined from the refinements. Fig. 3(b) compares the characteristic diffraction peaks of BSTS type chalcogenide[20] located at 2θ ~32.5º and 36.5º for MG, BG and MBG BSTS. These peaks are corresponding to the (107) and (00$\overline{12}$) crystal planes, respectively, where their appearances indicated the occupation of Se atoms in the center of the quintuple layer in our BSTS[20,27]. The peak intensities of their corresponding (107) and (00$\overline{12}$) crystal planes have been used to characterize the Bi(Sb)/Te anti-site and Se vacancy defects in the BSTS crystal[13,17,20]. The higher intensities of (107) and (00$\overline{12}$) diffraction peaks in MBG BSTS indicated the highly-ordered structure with relatively low composition-related defects compared to the MG and BG BSTS. The peak intensity ratios of (107) and (00$\overline{12}$) referred to (018), labeled as $I_{(00\overline{12})}/I_{(018)}$ and $I_{(107)}/I_{(018)}$, respectively, for the three-types of BSTS samples are compared quantitatively and plotted in Fig. 3(c).

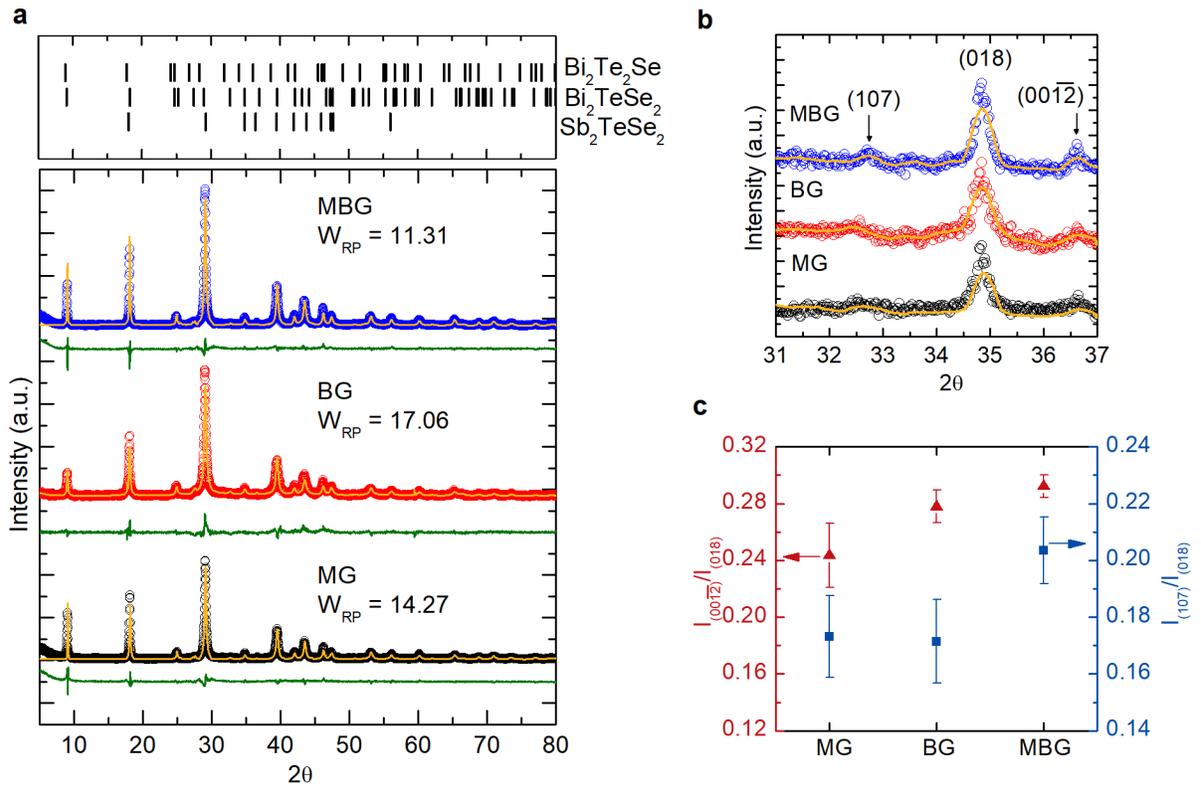

**Figure 3.** Powder XRD patterns of the BSTS samples grown by MG, BG and MBG methods (a). The extended view of the powder XRD patterns near the (108) peak to compare their characteristic

diffraction peaks (b). The $I_{(00\overline{12})}/I_{(018)}$ and $I_{(107)}/I_{(018)}$ of BSTS crystal grown by different methods (c).

**Electrical transport.** The electrical transport properties of the BSTS were studied to compare the quality of the crystals grown by different methods. Insets in Fig. 4(a) and (c) show the typical BSTS devices made by mechanical exfoliation and transfer onto pre-patterned electrodes on a Si/SiO$_2$ substrate. The gate-dependent four-probe resistances of the BSTS measured at room temperature (RT) and 1.5 K are compared for both MG and BG BSTS samples. Both BSTS samples display a broad resistance change over the gate voltage ($V_g$) range at RT. The MG BSTS shows a larger change at positive $V_g$ and nearly constant at negative $V_g$; while the BG BSTS reveals a clear ambipolar signature at RT. The broad resistance peaks indicate a combination of bulk and surface conductions. The bulk contribution is more significant in the hole-conduction (constant $R_{xx}$ region in negative $V_g$) in MG BSTS. As the samples cooled down to base temperature (1.5 K), a sharp resistance peak was revealed in both samples. The distinct resistance peak is attributed to the surface transport due to the Dirac dispersion nature of the topological surface states.

To further investigate the temperature dependent transport, the color maps of the $R_{xx}$ as a function of temperature (y-axis) and gate-voltage (x-axis) for MG and BG BSTS are shown in Fig. 4(b) and (d), respectively. For MG BSTS, the $R_{xx}$ increased as the temperature decreases, and the resistance peak maximized at ~30 K. This insulating behavior is attributed to the bulk dominating conduction due to its insulating nature (bulk gap). The insulating trend terminated at about 30 K, suggesting the suppression of the BSTS bulk conduction at the temperature[25]. A gradual decrease in $R_{xx}$ was observed by further reducing the temperature. As the surface states are gapless, this metallic behavior indicates prominently surface contribution to transport in this temperature range[25,28]. A similar temperature dependence profile was observed for the BG BSTS with resistance peak reaches maximum at ~50 K. Additionally, both devices showed a visible shift in resistance peak position with temperature near the region when the $R_{xx}$ reaching its maximum. This implied a shift of the chemical potential happened as the bulk conduction was suppressed and the total conduction developed into the surface conduction[28,29]. The origin of the resistance peak shifts is beyond the scope of the paper.

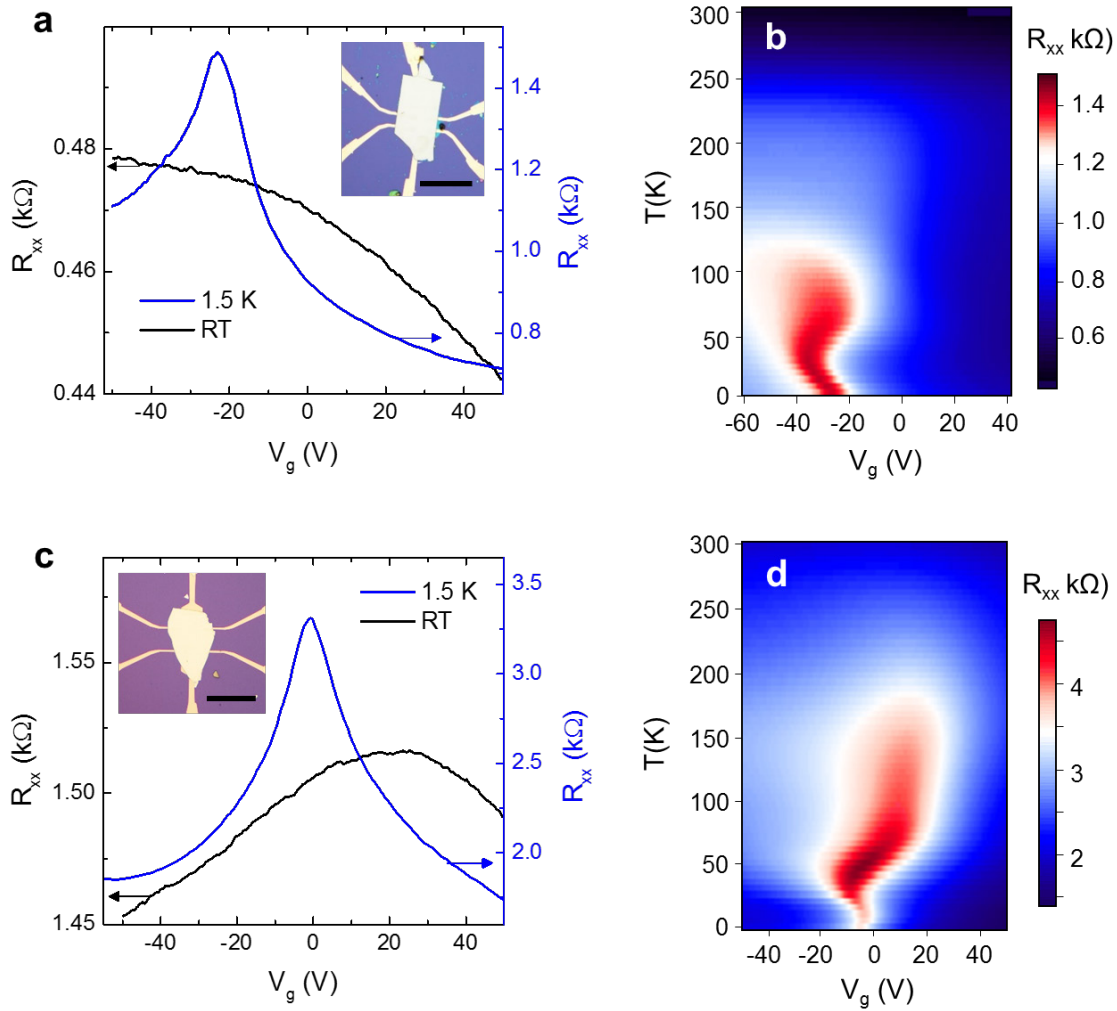

**Figure 4.** Gate-dependent resistances of MG (a) and BG (c) grown BSTS measured at RT and base temperature of 1.5 K. Optical images of the corresponding BSTS devices are inserted in (a) and (c) (scale bar= 50 μm). 2D color plots of the four-probe resistance as a function of temperature and gate voltage for MG (b) and BG (d) grown BSTS samples.

**Surface mobility study.** The low temperature magnetoelectric transport properties of three types of BSTS samples, grown by MG, BG, and MBG methods, are compared in Fig. 5. Fig. 5(a) shows the longitudinal resistivity ($\rho_{xx}$) of the MG, BG, and MBG BSTS devices as a function of gate voltage measured at base temperature (1.5 K) and zero magnetic field. Both BG and MBG BSTS reveal sharper resistivity peak and greater change in $\rho_{xx}$ compared to the MG BSTS. This indicates a better quality of the BSTS grown by vertical Bridgman furnace. The resistivity peak width of the MBG is about 33% narrower than the BG BSTS. This is consistent with the observation of the narrower peak width from the XRD analyses. The Hall resistivity ($\rho_{xy}$) plots of the BSTS devices as a function of gate voltage measured at 2 T are compared in Fig. 5(b). The sign change in $\rho_{xy}$ at the Dirac point confirms the ambipolar charge transport for all the BSTS devices. The $\rho_{xy}$ of MBG BSTS (~2.7 kΩ/T) and BG BSTS (~2.0 kΩ/T) develop much faster than MG BSTS (~0.3 kΩ/T) in magnetic field. The $\rho_{xy}$ of BG BSTS develops slower in the electron conduction region, while the MBG BSTS shows higher symmetry in both hole and electron conduction regions.

Hall mobility of the BSTS devices was calculated using the relation as: $\mu_H = \frac{\rho_{xy}}{\rho_{xx}} \frac{1}{B}$, from the linearly increasing region of $\rho_{xy}$ versus B (0–2 T). Fig. 5(c) compares the $\mu_H$ of the three BSTS devices at different charge density regions tuned by back-gate voltage. The $\mu_H$ of the MG BSTS maximized at ~1000 cm²/Vs in low electron density region. Both BG and MBG BSTS samples showed significantly higher $\mu_H$, which are about 3800 and 4400 cm²/Vs, respectively, in low hole density region. The mobility of our BG BSTS is comparable to values reported in literatures[23,25] (see Fig. S2 for the comparison of the surface mobility in literatures). However, the $\mu_H$ in the electron region of BG BSTS is about five times smaller than the hole conduction region. The electron mobility of the BG BSTS is markedly improved by recrystallizing the single crystal, as revealed by MBG BSTS. The extremely high surface mobility in MBG BSTS is attributed to the enhanced crystallinity of its parent crystal as identified by XRD.

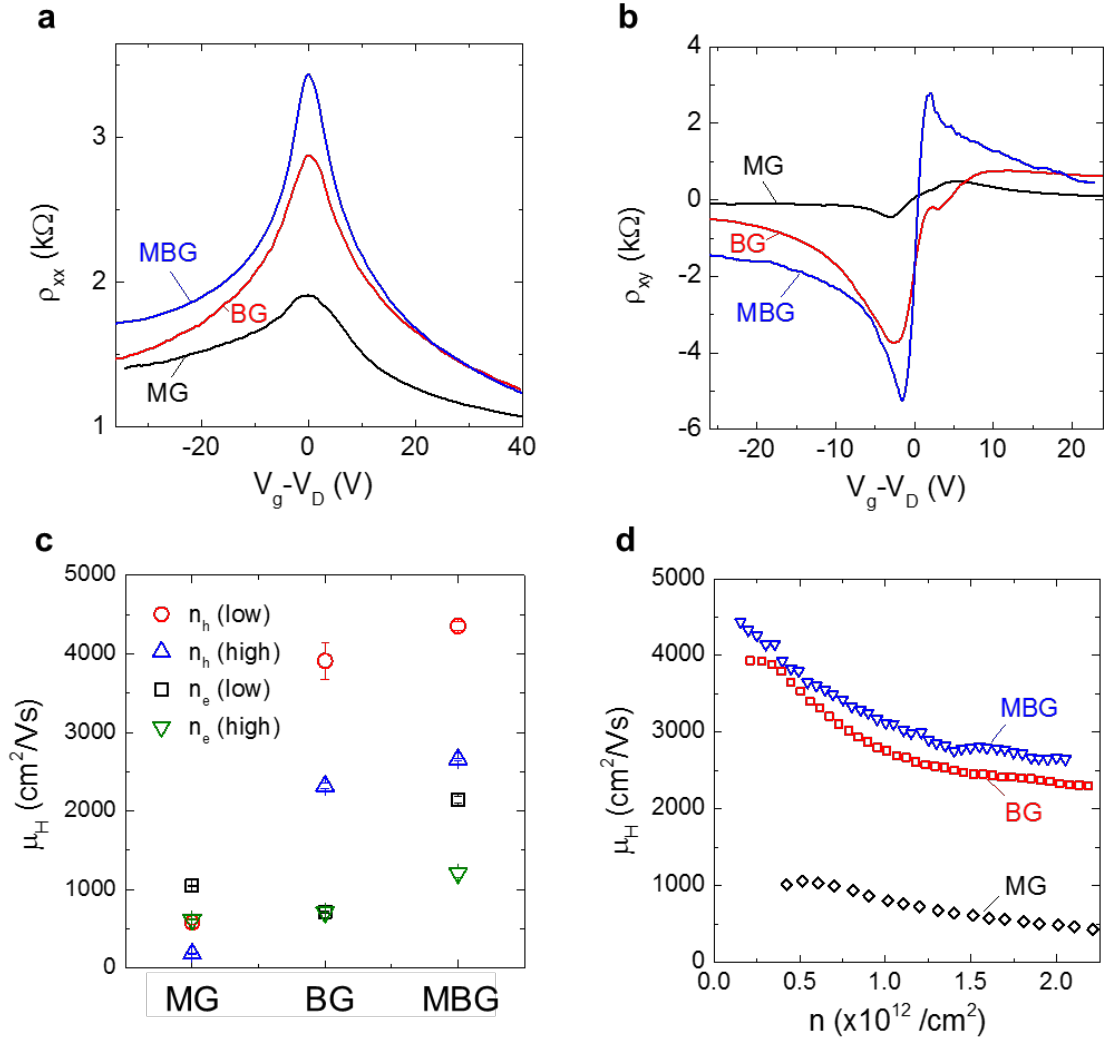

**Figure 5.** Longitudinal (a) and Hall (b) resistivities of the MG, BG and MBG BSTS devices as a function of gate voltage measured at magnetic field of 0 T (a) and 2 T (b). Comparison of the Hall
12

mobilities of the three-methods grown BSTS devices at different carrier densities at base temperature of 1.6 K (c). Hall mobilities as a function surface charge density for BSTS grown by three different methods (d).

**Quantum transport.** The quantum magneto-transport of the three BSTS devices are compared in Fig. 6. The Hall resistivity ($\rho_{yx}$), conductivity ($\sigma_{yx}$) and longitudinal resistivity ($\rho_{xx}$), conductivity ($\sigma_{xx}$) of the MG, BG and MBG BSTS as a function of magnetic field are presented in Fig. 6(a) and (b), respectively. For MG BSTS device, the $\rho_{yx}$ and $\rho_{xx}$ increase monotonically with magnetic field, which indicates that the surface states are in the normal Hall effect regime. For BG BSTS, the $\rho_{yx}$ approaches the quantum limit (25.8 k$\Omega$) of the integer quantum Hall effect (IQHE) at 9 T, together with the decreasing of $\rho_{xx}$ at magnetic field above 4 T. Meanwhile, the $\rho_{yx}$ of MBG BSTS reaches quantum Hall regime at magnetic field about 7 T as indicated by the saturation of $\rho_{yx}$ and the suppression in $\rho_{xx}$. The $\sigma_{yx}$ are plateauing at +1$e^2$/h at about 7 and 5 T for BG and MBG BSTS devices, respectively, as shown in Fig. 6(a). The corresponding $\sigma_{xx}$ (Fig. 6(b)) vanishes to <0.1 $e^2$/h. The development of IQHE in both BG and MBG BSTS suggests a better-quality BSTS crystal grown by Bridgman as compared to melting method.

To further confirm the IQHE in the BG and MBG BSTS, we investigate the quantization states by controlling the gate voltage. Fig. 6(c) and (d) present the $\sigma_{xy}$ and $\sigma_{xx}$ of the MG, BG and MBG BSTS as a function of gate voltage. The MG BSTS shows no sign of $\sigma_{xy}$ plateau formation at the any integer and the $\sigma_{xx}$ is far from vanishing (>2 $e^2$/h). On the other hand, the BG BSTS displays clear Landau level (LL) formation in $\sigma_{xy}$ with filling factor, $\nu$= -1, 0 and +1 by tuning the gate near the Dirac point. The LL states show different minima in $\sigma_{xx}$ as indicated by the arrows. The IQHE has been observed in BG BSTS by several groups[23-25]. The IQHE is explained as a combination of half-integer QHE arising from the top and bottom surface states. The LL filling factor (ν) can thus be expressed as, $\nu = \nu_t + \nu_b$, with $\nu_t = \left(n_t + \tfrac{1}{2}\right)$ and $\nu_b = \left(n_b + \tfrac{1}{2}\right)$, where $n_t$ and $n_b$ are the LL indices of the top and bottom surfaces, respectively. For the MBG BSTS with higher mobility, an additional bump at $\sigma_{xy} \sim 0.25\ e^2/h$ was observed in $\nu = 0$ quantum state. This could be a developing quantum state, which requires further studies to confirm its origin.

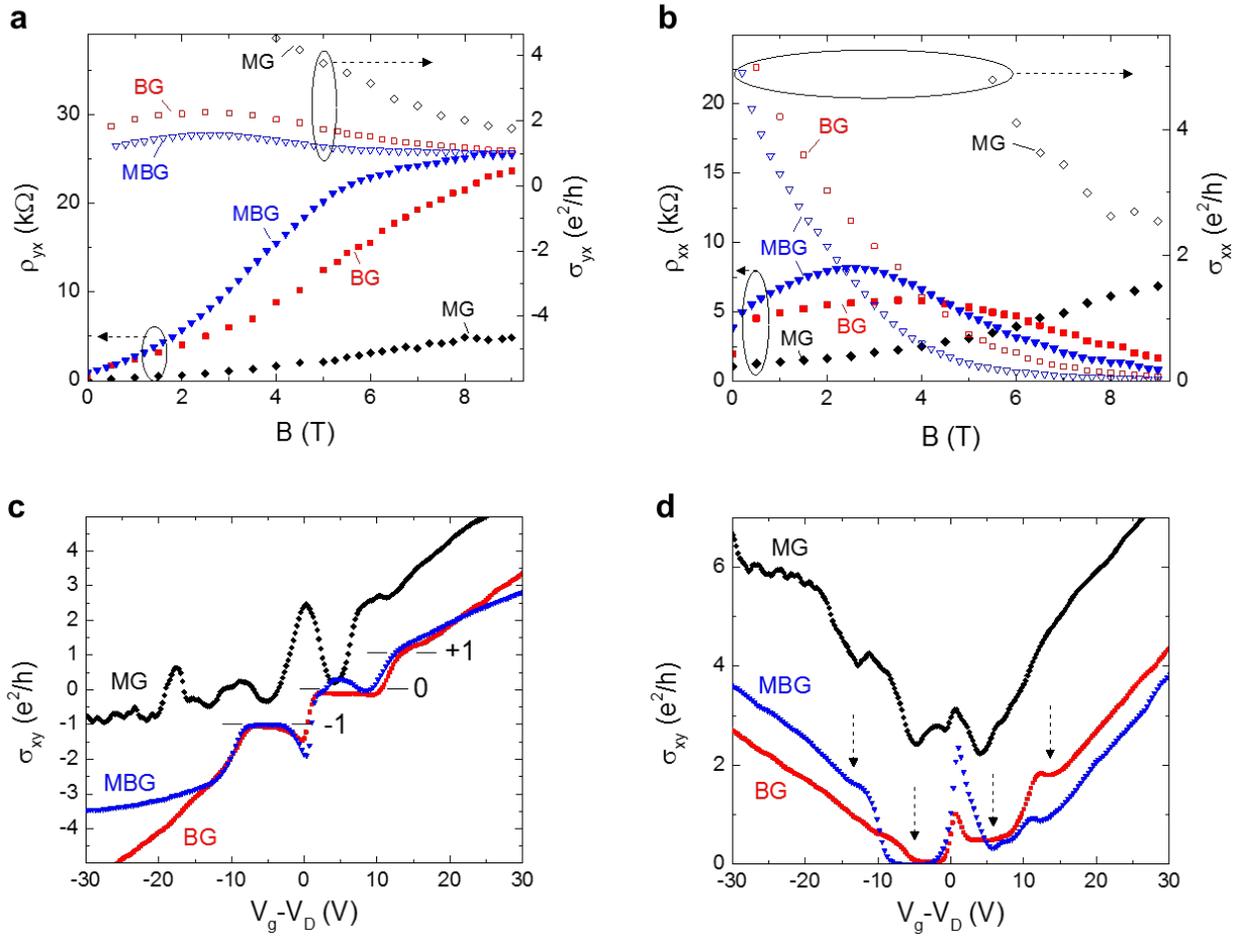

**Figure 6.** Plots of the Hall resistivity and conductivity (a), and the longitudinal resistivity and conductivity (b) as a function of magnetic field for the MG, BG and MBG BSTS devices. The Hall conductivity (c) and longitudinal conductivity (d) as a function of gate voltage at 9 T and 1.5 K for three methods grown BSTS devices.

## Discussion

Based on the observations, we showed that growth method has a strong effect on the crystal quality of the BSTS. We attributed this to the difference in growth mechanisms between the MG and BG methods. The MG sample solidifies from numerous nucleation sites along the wall of the ampoule. Whereas BG crystal nucleates from the bottom tip of the ampoule and solidifies along the vertical direction from bottom to the top. The multiple growth sites in MG are more likely to form grain boundaries when joining to form a bigger crystal. Crystal defects such as anti-sites, vacancies and other crystal disorder were also be reduced by the BG and MBG method. This minimized the electron scattering due to the defect sites in transport, and therefore resulted in a nearly four times enhancement in surface mobility of the BSTS. The enhancement in surface mobility led to the manifestation of IQHE in BG and MBG BSTS, in contrast to the MG BSTS.

Despite its high hole mobility, the electron mobility of the BG BSTS is relatively low (referred to Fig. 5(c)). We ascribed this to the imperfect composition stoichiometry of the BG BSTS, as revealed by the EDS and ICP-MS results. Tuning the composition of the four elements is challenging because the stoichiometry of $BiSbTeSe_2$ is not fixed thermodynamically (unlike $Bi_2Se_3$ or $Bi_2Te_2Se$) in phase diagram[30]. We argued that the two-step MBG method can resolve the inhomogeneous composition by recrystallizing the MG BSTS single crystal. The recrystallization in vertical Bridgman rearranged the molecules into ordered structure by directionally melting and solidifying, as indicated by the XRD patterns of the MBG BSTS. In addition, the quartz rod added on top of the single crystal ingot (made of the MG BSTS) minimized the headspace of the ampoule. This essentially limited the space for the vapor molecules to circulate in the ampoule, and therefore led to a composition more closely matching the desired stoichiometry[26]. As a result, we observed the mobility of electrons in MBG BSTS improved to >2000 cm$^2$/Vs as compared to BG BSTS (<1000 cm$^2$/Vs).

## Methods

**Ampoule preparation.** The metal trace of bismuth (Bi), antimony (Sb), tellurium (Te), and selenium (Se) (Sigma-Aldrich Co., purity 5N grade). The quartz tubes (Technical Glass Products, Inc.) with outer diameter of 14 mm and 1 mm thickness of wall were used for ampoule preparation. The raw materials were weighed to the molar ratio of 1:1:1:2 for Bi:Sb:Te:Se and mixed into a mixture with total weight of 5 g. The quartz tube was sealed one end into a conical shape. The inner wall of the tube was coated with an inert carbon layer via pyrolysis of acetone to prevent the reaction of the materials with the tube. The mixture in the quartz tube was flushed with argon gas a few times to displace the air out from the tube. This followed by an evacuation of the tube to a pressure of below 10$^{-6}$ torr to ensure a high vacuum environment for the crystal growth. The tube was then sealed by a torch into an ampoule with length of 60-80 mm.

**Melting growth.** Melting growth BSTS single crystals were prepared in a muffle furnace (F30438CM, Fisher Scientific, Waltham, MA). The ampoule was placed at the center of the muffle furnace. The temperature of the furnace was increased to 850°C at a slow rate of 0.1°C/min and was held for 48 h. A few times of intermittent mixing were performed by gently shaking the ampoule at the temperature. The ampoule was then cooled down to 550°C at a rate of 0.1°C/min, and was annealed at the temperature for 96 h. After that, the sample was slowly cooled down to room temperature at a rate of 0.1°C/min. The temperature as a function of time for the melting growth method was illustrated in Fig. 1(a).

**Bridgman growth.** Bridgman growth was carried out in a vertical Bridgman furnace with three coil heaters[31]. The ampoule was held by a thin string at one end and placed vertically at the level above first heating zone. While the other end of the string was attached to a motion controller. The temperatures were set to 670°C, 770°C and 500°C for warm, hot and cold zones, respectively, with the sequence from top to bottom heaters. The temperature profile was illustrated in Fig. 1(b). The first heater acted as a pre-heater to prevent the deposition of Se vapors on the wall of the ampoule[32]. The ampoule was translated vertically downward through the furnace at a very slow rate of 6 mm/day.

**Two-step melting and Bridgman growth.** The single crystal BSTS was first grown in the muffle furnace by following the steps discussed in melting growth process. The as-grown single crystal BSTS was placed in a new carbon coated quartz tube with a quartz rod placed on top of the sample. The second growth was carried out in the vertical Bridgman furnace in the same processes described in the Bridgman growth method.

**Samples characterization.** The as-grown crystals were cleaved and exfoliated along the crystalline surface for energy-dispersive X-ray spectroscopy (EDS) characterization. EDAX EDS (equipped in FEI Quanta 600 field emission scanning electron microscopy), operating at an acceleration voltage at 15 kV, was utilized to obtain EDS signals. Elemental compositions of the crystal were also studied using an inductively coupled plasma mass-spectrometry (ICP-MS, Agilent 7500 series, Agilent Technologies Inc.). For specimen preparation, the BSTS flakes were dissolved into a mixed nitric acid ($HNO_3$, 1.2 M) and hydrochloric acid (HCl, 0.3 M) solution. Bruker D2 Phaser X-ray diffractometer with Cu $K_\alpha$ radiation and a zero-background holder (G130706, MTI Co.) was employed for crystal plane and powder X-ray diffraction (XRD) data acquisition. The full-pattern Rietveld refinement was performed using GSAS-II software to determine the crystal structure.

**Transport measurements.** The sample flakes were exfoliated by using scotch tape on a polydimethylsiloxane (PDMS, Sigma-Aldrich Co.) substrate and transferred onto the pre-patterned gold leads in a Hall bar configuration. The transport measurements were carried out at various temperature from room temperature down to 1.5 kelvin, and magnetic field up to 9 tesla. The Hall measurements were performed by a SR830 DSP lock-in amplifier, operating at frequency of 17.777 hertz and the constant AC excitation current of 100 nA. The DC gate voltage was applied to the gate electrodes by a Keithley 2400 source measure unit.

## Acknowledgements

This work was supported by the NSF MRSEC program at the University of Utah under grant # DMR 1121252.

## Author contributions

TDS and VVD initiated the research, supervised the experimental design and data analysis. KBH performed the crystal growth and characterized the samples. SKC fabricated devices and performed the transport measurements. KBH and SKC wrote the manuscript. TDS and VVD revised the manuscript. AN and MAS advised the crystal growth and characterization processes. SP assisted the experiments in crystal growth.

## Additional information

**Supplementary Information** accompanies this paper at http://www.nature.com/scientificreports

**Competing financial interests:** The authors declare no competing financial interests.